\documentclass[preprint,12pt]{elsarticle}

\usepackage{amssymb}
\usepackage{subfiles}
\usepackage{booktabs}
\usepackage{graphicx}
\usepackage{float}
\usepackage[T1]{fontenc}
\usepackage{mathtools}
\usepackage{cleveref}
\usepackage{float}
\usepackage{adjustbox}
\usepackage{subcaption}

\usepackage[skip=0.5\baselineskip]{caption}

\journal{Computers \& Security}

\begin{document}

\begin{frontmatter}



\title{An Ensemble of Pre-trained Transformer Models For Imbalanced Multiclass Malware Classification}


\author[inst1]{Ferhat Demirkıran}

\affiliation[inst1]{organization={Cyber Security Gradaute Program},
            addressline={Kadir Has Unviersity}, 
            city={Istanbul},
            country={Turkey}}

\affiliation[inst3]{organization={Huawei R\&D Center},
            city={Istanbul},
            country={Turkey}}

\affiliation[inst2]{organization={Management Information System},
            addressline={Kadir Has University}, 
            city={Istanbul},
            country={Turkey}}
            
\author[inst3,inst2]{Aykut Çayır}
\author[inst2]{Uğur Ünal}
\author[inst2]{Hasan Dağ}

\begin{abstract}

Classification of malware families is crucial for a comprehensive understanding of how they can infect devices, computers, or systems. Hence, malware identification enables security researchers and incident responders to take precautions against malware and accelerate mitigation.
API call sequences made by malware are widely utilized features by machine and deep learning models for malware classification as these sequences represent the behavior of malware. However, traditional machine and deep learning models remain incapable of capturing sequence relationships among API calls. Unlike traditional machine and deep learning models, the transformer-based models process the sequences in whole and learn relationships among API calls due to multi-head attention mechanisms and positional embeddings.
Our experiments demonstrate that the Transformer model with one transformer block layer surpass the performance of the widely used base architecture, LSTM. Moreover,  BERT or CANINE, the pre-trained transformer models, outperforms in classifying highly imbalanced malware families according to evaluation metrics: F1-score and AUC score.
Furthermore, our proposed bagging-based random transformer forest (RTF) model, an ensemble of BERT or CANINE, reaches the state-of-the-art evaluation scores on the three out of four datasets, specifically it captures a state-of-the-art F1-score of 0.6149 on one of the commonly used benchmark dataset.


\end{abstract}




\end{frontmatter}


\section{Introduction}
\label{sec:sample1}

In recent times, with our dependence on information technologies, the Internet has been widely used by people of all ages. Those who want to quickly meet their daily needs such as online banking, online shopping, health, and transportation-related transactions cause an enormous increase in internet usage as well. This exponential growth of the usage of Internet plays a significant role in making life easier. On the other hand, this situation poses a severe threat as cyber attacks increase drastically in parallel with the growth of the Internet. Among these cyber attacks, malicious software (malware) is the primary weapon for attackers to conduct their malicious activities against a victim's machine such as computer, smartphone, or computer networks in order to disrupt system's functions and gain unauthorized access \cite{jang2014survey,aslan2020comprehensive}.

Cybercriminals use several ways to spread malware, such as phishing e-mails with malicious links and attachments, text messages, and malicious advertisements etc. According to the state of e-mail security report released by Mimecast in 2021, 61\% of organizations were exposed to e-mail-based ransomware in 2020, with an increase of 10\% compared to the previous year \cite{mimecast}. The average amount spent to recover from a ransomware attack, when factors such as downtime, device, human, and network costs are included, is about \$1.85 million as reported by Sophos \cite{sophos}.
According to another cyber threat report published by SonicWall, 5.6 billion malware attacks were carried out in 2020 \cite{sonicwall}. 


In lieu all of these findings, one can safely claim that excessive malware, without considering the identification/classification methods, affects many victims destructively. Since the numbers of malicious software and the damages they cause to the institutions are increasing every day, it is crucial to map malware behavior that can be provided by malware family identification so that security researchers and incident responders can speed up the recognition and mitigation processes.


There are two main approaches used the most to detect malware. One of them is the signature-based malware detection method. The signatures, sequences of bytes, created using static, dynamic, or hybrid methods are uniquely located in the database. Whether a given file is malware or not is determined by looking at the unique signature of this file from a predefined database \cite{shijo2015integrated}. Although signature-based methods are the most generally utilized procedure in antivirus programming, since the only predefined list of known malware variants are kept, they are not able to catch previously unidentified malware \cite{ucci2019survey}.

The other main malware detection approach is behavior-based method, which  examines the behavior and characteristics of a given file and then decides whether the related file is malware, and if it is a malware, then  the approach also defines the malware family the file belongs to.
Although the effort and time spent and the storage complexity are much more, the unknown attacks can be detected and classified by using behavior-based methodologies better contrary to the signature-based methods \cite{gibert2020rise}.

In the report released by SonicWall, among detected malware, 268,362 of them have never been seen before in 2020, with a rise of 74\% from the preceding year \cite{sonicwall}. Considering the increasing number of unseen malware over the years, performing a behavior-based approach is more reasonable. This report indicates the significance of developing more innovative and effective malware defense mechanisms to detect and classify unknown malware. 


The effectiveness of the malware defense mechanism is directly associated with the right choice of behavioral features exploited from malware. Several features can be extracted from malware due to its diverse nature. Obtaining adequate  features is time-consuming for a model. This situation can make learning difficult for a model if some of the features used are non-distinctive \cite{jindal2019neurlux}. In our study, both static and dynamic API call sequences are leveraged to classify malware families since these sequences represent behavioral patterns for each sample. Considering API call sequences, machine learning becomes the primary choice to capture sequence relationships between the sequence elements for malware classification. 



Different machine learning algorithms have been used in the literature for malware detection and classification so far \cite{komatwar2021survey,ucci2019survey}. Considering the sequence, traditional machine learning models may not be sufficient as the relations among API calls must be taken into account to successfully predict the malware families of unseen API call sequences.
The current deep learning based models, mainly pre-trained transformer models outperform traditional machine learning based approaches for sequential text classification  \cite{li2020survey,minaee2021deep}. 



In this paper, we have answered the following research questions respectively:

\begin{quote}\emph{RQ.1: What are the suitable classification metrics for imbalanced datasets in multiclass malware classification?}\end{quote}

\begin{quote}\emph{RQ.2: What are the appropriate base models for multiclass malware classification based on API call sequences?}\end{quote}

\begin{quote}\emph{RQ.3: What are the effects of pre-processing on API call sequences to the model results?}\end{quote}

\begin{quote}\emph{RQ.4: What are the effects of tokenizer-based (word piece) pre-trained transformer model (e.g. BERT) and tokenizer-free transformer model (e.g. CANINE) to our model results?}\end{quote}


\begin{quote}\emph{RQ.5: What is the effect of ensemble of pre-trained transformer models, BERT and CANINE, which is based on bagging for imbalanced multiclass malware classification?}\end{quote}







Our main contributions through this study, in the light of the answers to our research questions, can be summarized as follows:

\begin{itemize}

\item Noticing inconsistent evaluation results due to a logical error in the code of a published article \cite{schofield2021comparison}.

\item 
To the best of our knowledge, we have used the pre-trained CANINE transformer model for the first time in the field of malware in this study.

\item Again, to the best of our knowledge, a bagging-based ensemble of pre-trained transformer models has been used for the first time in malware classification.

\item Our proposed model Random Transformer Forest (RTF), has surpassed the state-of-the-art results obtained in the malware classification.

\item We have achieved a state-of-the-art result on one of the well-known API call dataset in the literature \cite{catak2019benchmark} with our proposed RTF model.

\end{itemize} 

This paper is structured as follows: Section \ref{sec:related_work} presents the related work. The description of the datasets, base models, pre-trained models, and our proposed model are presented in Section \ref{sec:methodology}. The test results are discussed and compared with the related studies in Section \ref{sec:exp_res}, and lastly, the conclusion and future work are given in Section \ref{sec:conclusion}.



\section{Related Work}
\label{sec:related_work}
Cybercriminals leverage malware to exploit any device or system to steal sensitive data and hence cause enormous problems for victims. Analyzing and classifying incoming malware helps us define the problem and understand how to recover from the damage as quickly as possible.

There are two techniques most commonly used in malware analysis, static analysis and dynamic analysis. Static analysis is a process of malware analysis that analyzes the given malware without running it. Unlike static analysis, a given malware file is executed in an isolated environment to avoid harming the computer system in dynamic analysis.  

Malware developers may implement various techniques to evade detection mechanisms such as code obfuscation, dynamic code loading, polymorphism, and metamorphism. For instance, the MD5 hash based detection method can be easily bypassed by malware authors with the methods mentioned above. As these methods cause the binary of the file to change, they also cause a change in the hash of the file. While the hash of the malicious file is changed and the file is defined as benign, the behavior of the file, thus its effect, remains unchanged \cite{ucci2019survey}.

Since dynamic analysis requires the execution of a given sample to be monitored and observed in an isolated environment, malware even written with code obfuscation techniques hardly eludes dynamic analysis contrary to static analysis. This identified situation provides dynamic analysis to be more robust than static analysis \cite{or2019dynamic}. 


Performing dynamic analysis requires more time than static analysis and organizations are dealing with millions of attacks carried out in a day. 
These shortcomings provide an excellent opportunity for machine learning to collaborate with dynamic and static analysis since machine learning can handle large volumes of data \cite{fraley2017promise}. 

In the malware detection and classification process, understanding malware behavior is one of the substantial parts of detecting and classifying malware. Dynamic API calls are obtained by tracing the sequences of calls by way of calling operating system services such as creating a file and allocation of virtual memory by malware samples. On the other hand, static API calls are extracted from portable executable (PE) format of the executable files. Static API calls are unordered as the sequences of calls are not traced unlike Dynamic API calls \cite{han2019maldae}. In general, since API call sequences generate specific behavioral patterns and hence represent malware families, they can be considered as one of the most distinguished features among malware families \cite{ding2018malware,fujino2015discovering}.

Related studies about base models for malware analysis using API calls and transformer-based models on sequence problems will be examined respectively in the rest of this section. 

\subsection{Base Models for Malware Analysis using API Calls}


In \cite{ki2015novel}, the authors  used DNA sequence alignment algorithms, Multiple Sequence Alignment (MSA), and Longest Common Subsequence (LCS) to extract the most critical API call sequence patterns among different malware families and generate a signature-based malware detection mechanism to determine whether a program is a malware or not based on these extracted patterns. The API call sequences determined by the MSA and LCS algorithms can be misleading for the model if sequences get more extended than a preset API call sequence length. 

In \cite{sundarkumar2015malware}, the authors proposed a model using text mining and topic modeling for feature extraction and selection processes based on API call sequences. Machine Learning based Group Method of Data Handling (GMDH) method, traditional machine learning models, Random Forest (RF), Decision Tree (DT), Support Vector Machine (SVM), and Multilayer Perceptron (MLP) are compared on two different datasets. Although DT and SVM models outperformed the results, and they suggest DT for malware detection expert system, the size of the datasets are inadequate to rely on the models.  

In \cite{kolosnjaji2016deep}, the authors integrated Convolutional Neural Network (CNN) and Recurrent Neural Network (RNN) layer into one neural network architecture. With this model, they have achieved the accuracy score of 89.4\%, the precision score of 85.6\%, and the recall score of 89.4\% to classify malware families. Newly generated subsequences of original API call sequences are given to the model as an input. For each API call sequence, if the same API call is repeated more than two times in a row, only two consecutive identical API calls are included in the resulting sequence. Since their corpus contains only 60 different API calls, they did not set any boundaries. Otherwise, they may have to set a predetermined length to avoid tracking loops and make the model less complex. 

In \cite{tobiyama2016malware}, two stages of Deep Neural Networks are applied for the malware detection process. The proposed model CNN is used to classify feature images extracted with Long Short-Term Memory (LSTM) model using API calls. Although the authors achieved an Area Under Curve (AUC) score of 96\%, since the size of the dataset is relatively small, the score may be misleading.

In \cite{mathew2018api}, the authors leveraged N-gram and Term Frequency–Inverse Document Frequency (TF–IDF) for feature extraction and selection, respectively. The proposed LSTM model is used for binary classification, benign or malware, using API call sequences. The authors reached a 92\% accuracy score on unknown test API call sequences.

In \cite{xiao2019android}, the authors trained two different LSTM networks on system call sequences for both malware and benign Android applications, respectively. The new sequence has been classified by comparing two similarity scores obtained from two different LSTM networks. The LSTM model has been compared with two n-gram models, MALINE and BMSCS, based on accuracy, precision, recall, and False Positive Rate (FPR). They have shown that the LSTM model outperformed MALINE and BMSCS models.   

In \cite{catak2020deep}, the authors tried several models from LSTM to traditional machine learning models, RF, DT, SVM, and K-Nearest Neighbour (KNN), on a dataset containing 7,107 samples of API call sequences generated by them. In multiclass classification, they have achieved the highest F1-score of 47\% using the single-layer LSTM model compared to the two-tier LSTM and common machine learning models. 

In \cite{li2021api}, the authors generated new API call sequences by applying data pre-processing steps to benchmark dataset \cite{catak2019benchmark}. If any unique API call repeated more than once in the sequence, they kept only one and removed the continuously same API.  Similarly, they removed continuously same sub sequences when the length of sub sequences are two or three. They have proposed two different models, LSTM and Gated Recurrent Unit (GRU) models, which are based on RNN architecture. They have significantly increased their precision, recall, and F1-score after data pre-processing.

In \cite{oliveira2019behavioral}, similar to feature pre-processing method applied in previous work \cite{kolosnjaji2016deep}, the authors prepared new API call sequences and kept only 100 non-consecutive sequences to avoid repeating API calls loop. They achieved similar AUC score and F1-score results compared to LSTM with their proposed model, which is based on Deep Graph Convolutional Neural Networks (DGCNNs). 



LSTM model is widely used as the underlying architecture for malware detection and classification based on API calls, as seen in the above mentioned studies.
\subsection{Transformer Based Models}

This section is structured as follows: Section \ref{sec:PTM} introduces the pre-trained transformer models. Section \ref{sec:TBM} presents the related work relevant to the transformer based models on sequence problems.

\subsubsection{Pre-trained Transformer Models}
\label{sec:PTM}

With the increasing use of deep learning approaches, the number of model parameters and hence the need for a much larger dataset to train these model parameters have increased. Since constructing a large labeled dataset is time-consuming and requires extremely expensive annotation costs, contrary to constructing a large unlabeled dataset, pre-trained models have gained importance. Models that are pre-trained on the huge unlabeled text data learn good universal representations, which are used to fine-tune the model on the downstream tasks \cite{qiu2020pre}. On the other hand, the pre-trained transformer models have reached state-of-the-art results as these models are capable of capturing dependencies over a wide range of scales unlike convolutional and recurrent networks \cite{ganesh2021compressing,lin2021survey}. Pre-trained techniques leveraged to capture these dependencies are elaborated in the sections \ref{sec:Bert} and \ref{sec:Canine}.


\subsubsection{Transformer Based Models on Sequence Problems}
\label{sec:TBM}

In \cite{erciyes2021deep}, the authors used several deep learning models from mostly used traditional deep learning methods such as CNN, RNN, LSTM, GRU, and BiLSTM with GLOVE and fastText embedding to pre-trained transformer models for multiclass text classification. For their experiment, they utilized the highly imbalanced RCV1-v2 dataset, which contains 800,000 news stories. They have shown that transformer models outperformed traditional deep learning models based on F1-score for multiclass text classification.

In \cite{paul2020cyberbert}, the authors used the BERT model for the cyber-bullying detection task. The proposed model has been tested on three different datasets taken from Twitter, Wikipedia, and FormSpring. Compared to common machine learning models, SVM and Logistic Regression (LR), and deep-learning based models, CNN, RNN + LSTM, and Bidirectional LSTM (BiLSTM), they have achieved higher F1-scores. 

In \cite{alvares2021malware}, the authors generated word embeddings for each opcode of malware samples by using Word2Vec and BERT. They classified malware with different classifiers such as LR, SVM, and MLP to see the effect of different word embeddings. They have achieved higher results using BERT for word embeddings with the same set of input parameters and the same set of classifiers based on classification accuracy among five unique malware families distributed almost balanced.

In \cite{nassar2021malware}, they have proposed transformer-based architecture for detection and classification of malware using opcode sequences of windows executable files. The proposed transformer model has achieved better results compared to Gradient Boosting Method (GBM) and BiLSTM based on accuracy, precision, recall, and F1-score evaluation metrics. 


In \cite{xu2021malbert}, the authors proposed a pre-trained transformer model, Malbert, which is pre-trained on 15,000 malware and 15,000 benign samples \cite{ki2015novel} first to learn the relationships among API calls. This pre-trained transformer model and existing pre-trained transformer model, \emph{Bert-base-uncased} are fine-tuned on two different datasets for the malware detection process. Pre-trained transformer models have achieved higher results compared to LSTM model and traditional machine learning models based on mostly used evaluation metrics, such as accuracy, precision, recall, and F1-score. 

In \cite{mcdonnell2021cyberbert}, the authors proposed a model, called CyberBert which uses bidirectional transformer architecture for two different tasks, session-based recommendation, and malware classification based on API calls. Compared to the LSTM model and transformer-encoder,  a unidirectional model,  they have achieved higher F1-scores for binary and multiclass classification with CyberBert.

In \cite{oak2019malware}, the authors leveraged pre-trained BERT transformer model for malware detection, malware, and benign, on Android operating system API calls called by the application. The set of experiments made by the study shows BERT model obtained state-of-the-art results compared to the LSTM model on sequence classification.


Recent surveys \cite{qiu2020pre,minaee2021deep} with the studies mentioned above clearly show that current transformer-based models, mainly pre-trained transformer models fine-tuned on downstream tasks, outperformed traditional machine and deep learning models on sequence classification. 

\section{Methodology}
\label{sec:methodology}

In the methodology section, firstly, the datasets used in experiments are introduced. Secondly, the most suitable evaluation metrics for highly imbalanced datasets are specified. Then, base model structures, the effect of the pre-processing method, pre-trained transformer models, CANINE and BERT, and the proposed RTF model architectures are explained. 

\subsection{Datasets}

To verify how effectively a model classifies malware, it is necessary to test the model on different malware datasets. Since malware constantly evolves, working on an up-to-date malware dataset is required to assess the effectiveness of proposed models. Comparing several models on a single dataset containing outdated malware samples and highlighting one model might not be reliable. 
Thus, four different datasets containing API call sequences of malware samples and their corresponding malware families are utilized to evaluate the models used in this study. Of these four datasets used in this study, Catak and Oliveira datasets are created with dynamic API calls and VirusSample, whereas VirusShare datasets are constructed with static API calls.
\subsubsection{Dynamic API Call Datasets}

\paragraph{Catak Dataset} \mbox{}\\
This study obtained sequences of Windows Operating System API calls within the Cuckoo Sandbox isolated environment for each malware file. Malware family labels were determined using unique hash codes of each malware on the Virus Total website. In total, 7,107 samples, which contain hash codes of malware,  Windows operating system API call sequences, and their malware family classes, were created \cite{catak2019benchmark}.


\paragraph{Oliveira Dataset}\mbox{}\\
42,797 malware and 1,079 benign API call sequences were obtained via Cuckoo Sandbox for dynamic malware analysis. Instead of using whole API call sequences, the first 100 non-consecutive API call sequences were extracted from the parent processes to reduce complexity and detect the malicious pattern as quickly as possible. The generated dataset containing hashcodes, label (malware or benign), and 100 non-consecutive API calls for each sample has been used for binary malware classification \cite{oliveira2019behavioral}. 

Since we are working on a multiclass classification problem, malware families of 42,797 malware samples are determined through virus total. Out of 42,797 malware samples, 2,081 were labeled as "unknown" by virus total. Several malware families hold a small number of malware samples, less than 100. These malware samples are removed since they could be misleading for the models. Thus, the dataset in question has been turned into a multiclass classification case. The compiled dataset used in our study consists of 40,566 malware samples with their API call sequences and malware families.




\subsubsection{Static API Call Datasets}

\paragraph{VirusShare}\mbox{}\\
Unique hash codes represent malware samples obtained from Virus Share. Each unique hash code in text files is passed to Virus Total to learn their corresponding malware families. Python module named PEfile is leveraged to extract API calls from each malware sample. Lastly, malware families having less than 100 samples are removed.
Thus, 13,849 malware samples with their corresponding API call sequences and malware families are obtained \cite{duzgun2021new}.


\paragraph{VirusSample}\mbox{}\\
Malware samples taken by Virus Sample are kept with their unique hash code text files. Corresponding malware families and API calls are obtained from Virus Total site and PEfile module, respectively, as in Virus Share. Finally, malware families having less than 100 are removed from the dataset. 
Therefore, 9,732 malware samples with their corresponding API call sequences and malware families are obtained \cite{duzgun2021new}.
Since the malware samples in this dataset consist of the most up-to-date data based on API calls, we also find an opportunity to test our models on recent malware samples. Table \ref{table:classDistribution} shows the total samples of malware families for each dataset.

\begin{table}[ht]
\centering
\caption{Distribution of malware families.}
\begin{tabular}{@{}ccccc@{}}
\toprule
\multicolumn{1}{c}{\textbf{Malware Family}} & \textbf{Oliveira } & \textbf{VirusShare } & \textbf{Catak } & \textbf{VirusSample } \\ \midrule
Trojan                                     & 31,979	& 8,919		& 1,001          & 6,153			     \\
Virus                                  & 102         & 2,490& 1,001         & 2,367				   	       \\
Adware                                  & 5,444   & 908	& 379          & 222			      \\
Backdoor                               & 135         & 510	& 1,001          & 447		                    \\
Downloader 		         & 1,948  & 218	& 1,001          & N/A			     \\
Worms                                  &  N/A  & 524& 1,001          & 441				   	       \\
Agent                                     & 220& 	165	& N/A          & 102		       \\ 
Ransomware                         & 404&	115	& N/A          & N/A		   	       \\
Dropper                                   & 118      & N/A	& 891         & N/A			       \\
Riskware                                & 216&     N/A & N/A          & N/A				      \\
Spyware 			& N/A          & N/A	& 832          & N/A		   	       \\ \midrule
\textbf{Total}                              & \textbf{40,566}     & \textbf{13,849}             & \textbf{7,107}             & \textbf{9,732}                     
\end{tabular}
\label{table:classDistribution}
\end{table}

\subsection{RQ.1-) What are the suitable classification metrics for imbalanced datasets in multiclass malware classification?}

The degree of imbalance may vary within different domains. One of these domains is malware, as specific malware families are used chiefly in particular periods for cyber attacks.

According to the report released by Malwarebytes:
The total number of Trojan detected by Malwarebytes is almost 26 times higher than the total number of Worm in 2018. The total number of Riskware detected by Riskware tools in 2019 was 6,632,817, with a decrease of 35\% compared to the previous year.
In another chart containing the number of detection of malware families by months, it is seen that the number of Trojan attacks increased dramatically at the beginning of 2019, with the spread of the Emotet, one of the advanced Trojan campaign in that period \cite{malwarebytes}.

These situations demonstrate that there could be significant differences in the distribution of malware families according to years or even months. Therefore, when the collected malware is classified according to their families, the distribution will vary according to the malware type prevailing at the time of collection and hence lead to imbalance. 

For these reasons, almost all of the datasets belonging to malware have an imbalanced class distribution in the literature. Thus, we are required to leverage the most suitable metrics to evaluate our classification performance on imbalanced datasets.

The datasets leveraged in our study have highly imbalanced class distribution as expected and shown in Table \ref{table:classDistribution}.

Evaluation metrics are one of the crucial steps to assess model performance. An incorrectly chosen evaluation metric can make a poor performance algorithm seems effective. The metrics used to evaluate a model performance may vary for balanced and imbalanced datasets. For instance, using accuracy metric for a balanced dataset may provide an objective evaluation, yet may not be the right choice for an imbalanced dataset as it has a bias against the majority class. Taking malware classification for example. Assume there are six different classes in the dataset, and 95\% of the samples belong to Trojan. In this case, a dummy model that predicts all samples in the test data as Trojan will achieve an accuracy score of 95\% even though it does not predict any other classes correctly. It may not always be correct to use the most widely preferred evaluation metric without examining the distribution of classes in the dataset for the reasons mentioned above.

Recently, researchers have started to use Matthew's Correlation Coefficient (MCC) to evaluate model performance on imbalanced datasets. Although this metric was previously used mostly in biomedical research, it has now been used in many areas, including malware classifications \cite{kim2021machine,jahromi2020improved} yet according to the experimental results to investigate the behavior of MCC metric, MCC is not suitable for directly applying on imbalanced datasets \cite{zhu2020performance}. Empirical research conducted on 54 imbalanced datasets demonstrates that the AUC score is more discriminating than MCC \cite{halimu2019empirical}.
Most frequently used metrics to assess model performance for multiclass classification tasks have been shown to be inadequate on imbalanced datasets such as Precision, Recall, MCC, Confusion Entropy (CEN), Classes Average Accuracy (AvAcc), and Class Balanced Accuracy (CBA) \cite{branco2017relevance}.

Although the choice of right metrics is still an open issue, following the searches to find the most suitable metrics used for multiclass classification on imbalanced datasets, we have used AUC, which is a summary of probability curve, Receiver Operating Characteristic (ROC), based on FPR and True Positive Rate (TPR), as an evaluation metric. On the other hand, F1-score has been used to be comparable with studies conducted on one of the well-known datasets in the literature \cite{catak2020deep}. 

The equation (\ref{eqn:recall}) and (\ref{eqn:precision}) define the recall and precision metrics respectively.
The equation (\ref{eqn:f1score}) defines F1-score in terms of precision and recall. Also, the equation (\ref{eqn:f1score}) contains the explicit form of the formula in terms of True Positive (TP), False Negative (FN), and False Positive (FP).  

\begin{equation}
\label{eqn:recall}
Recall = \frac{TP}{TP+FN}
\end{equation}
\begin{equation}
\label{eqn:precision}
Precision = \frac{TP}{TP+FP}
\end{equation}
\begin{equation}
\label{eqn:f1score}
F_{1} score = \frac{2*Precision*Recall}{Precision+Recall} = \frac{2*TP}{2*TP+FP+FN}
\end{equation}



\subsection{Base Models}

LSTM and One-layer Transformer Block Based Transformer architecture are leveraged as base models for malware classification based on \emph{RQ.2}.
For the rest of the paper, One-layer Transformer Block Based Transformer architecture is referred to as \emph{Transformer} model. 


\subsubsection{LSTM Based Malware Classification}




Recurrent Neural Networks (RNNs) have a structure that uses recurrent relation, which is a situation of performing the same step, processing current output depends on the previously computed hidden state, for each element of a sequence repeatedly. In these structures, information is retained through the previous hidden state, and hence processing continues in time steps. The recursion between sequence elements hinders parallelization during the training phase and consequently causes a longer runtime for training. 

LSTM can learn relatively long-term dependencies compared to other RNNs because it provides deeper processing of hidden states through specific units. This situation causes an increase in the number of parameters used for training. Besides, since LSTM has a recursive structure like other RNNs inherently and hence can not be trained in parallel, the training period may take relatively longer compared to other RNNs \cite{hochreiter1997long}.

Since our purpose is to classify malware families, fully connected layer output that captured the information from the LSTM networks, is given to the softmax layer for multiclass classification. 

The standard LSTM network is preferred as one of the base models for comparison since it has been used widely as a base network and performed successfully for several malware classification problems using API call sequences \cite{berman2019survey}.









\subsubsection{Transformer Model}

Transformer model is a recently used network architecture that designed to overcome the deficiencies of sequence-to-sequence neural network approaches such as LSTM and RNN for the sequence modeling and transduction problems in 2017 \cite{vaswani2017attention}.

\begin{figure}[H]
\centering
\includegraphics[scale=0.60]{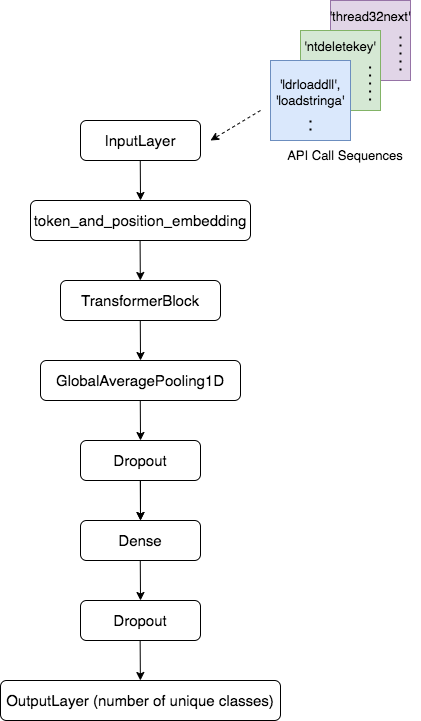}
\caption{Transformer model architecture.}
\label{fig:TransformerArchitecture.png}
\end{figure}

Since transformer-based architectures avoid recursion, they can overcome the parallelization problem that both the LSTM and other RNNs suffer from. 
In traditional sequence-to-sequence architectures the information coming from the previously hidden state is processed recursively to capture dependencies. On the contrary, since the transformer models refrain from recurrence and convolution, positional encodings are used to preserve the order of the sequence and provide position-related information of the tokens in the sequence. Transformer models leverage the attention mechanism to caption and preserve long-term dependencies for the sequences processed as a whole. Positional information is retained with attention layers instead of the recurrent and convolutional layers in transformer model \cite{vaswani2017attention}. Figure \ref{fig:TransformerArchitecture.png} shows the Transformer model architecture utilized in this study.



\subsection{Pre-processing Method on Datasets}

Each API call sequence is pre-processed, similar to the steps taken part in \cite{li2021api}. Pre-processing part consists of 3 main steps.

In the first step, for any API call in a sequence that repeated more than one time in a row, the continuously same API calls are removed from the sequence.
This pre-processing step generates a new sequence that does not consist of the consecutive same API call. 
In the second and third steps respectively, repetitive binary and triple sub-sequences are removed from the new sequence created by the first step. The following Figure \ref{fig:PYRAMID.png} shows the pre-processing step outputs respectively based on randomly given sequences.



\begin{figure}[H]
\centering
\includegraphics[scale=0.60]{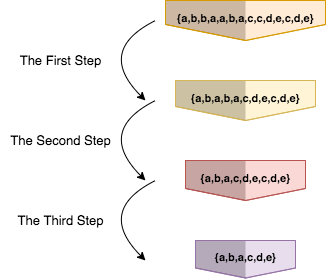}
\caption{The outputs of pre-processing steps.}
\label{fig:PYRAMID.png}
\end{figure}

These pre-processing steps have not been applied to the Oliveira dataset as this dataset has already given pre-processed and limited to the 100 non-consecutive API calls only. 
On the other hand, although the pre-processing steps have been applied to the VirusSample and VirusShare datasets as well, only one sample out of 9,732 samples for VirusSample dataset and only two samples out of 13,849 samples for VirusShare dataset are affected. Of these affected samples, only two or three API calls are affected. The results of the pre-processing steps applied on VirusSample and VirusShare datasets show us that static API calls do not include noisy and redundant API calls as expected contrary to dynamic API calls. Thus, we have continued with the original API call sequences of VirusSample and VirusShare datasets.
After performing pre-processing steps on Catak dataset, only 11 API call sequences remained constant.  
The effect of the pre-processing steps on the Catak dataset can easily be seen by the Figure \ref{fig:LengthDistributionOfEachApiCallSequences.png} shown below. 

\begin{figure}[H]
\centering
\includegraphics[width=\textwidth]{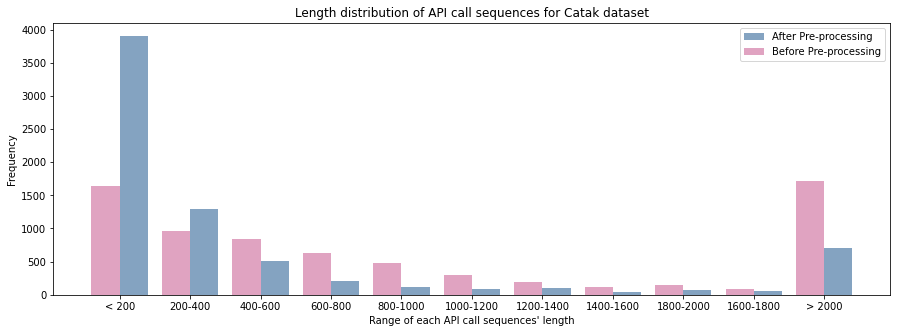}
\caption{The effect of pre-processing on Catak dataset.}
\label{fig:LengthDistributionOfEachApiCallSequences.png}
\end{figure}

After pre-processing steps, the number of samples where the length of the API call sequences is less than 200 has increased more than twice, and the number of samples where the length of the API call sequences is more than 200 has decreased more than twice. These changes clearly show the effect of pre-processing steps on the Catak dataset. 
Finally, after performing all the pre-processing steps to the Catak set only, the effect of the pre-processing is examined with model performances.

\subsection{CANINE and BERT}

Large-scale pre-trained models have recently become very popular in the field of artificial intelligence.
Due to the model previously trained on large-scale data, captured information can be used for specific tasks that utilize the pre-trained model by fine-tuning \cite{han2021pre}. 
This study utilizes two different pre-trained models architectures, BERT and CANINE.  

\subsubsection{BERT}
\label{sec:Bert}

BERT, Bidirectional Encoder Representations from Transformers, is a language model that uses transformer architecture which is pre-trained on Wikipedia and Book Corpus of unlabelled text \cite{devlin2018bert}. 

\begin{figure}[H]
\centering
\includegraphics[width=\textwidth,height=\textheight,keepaspectratio]{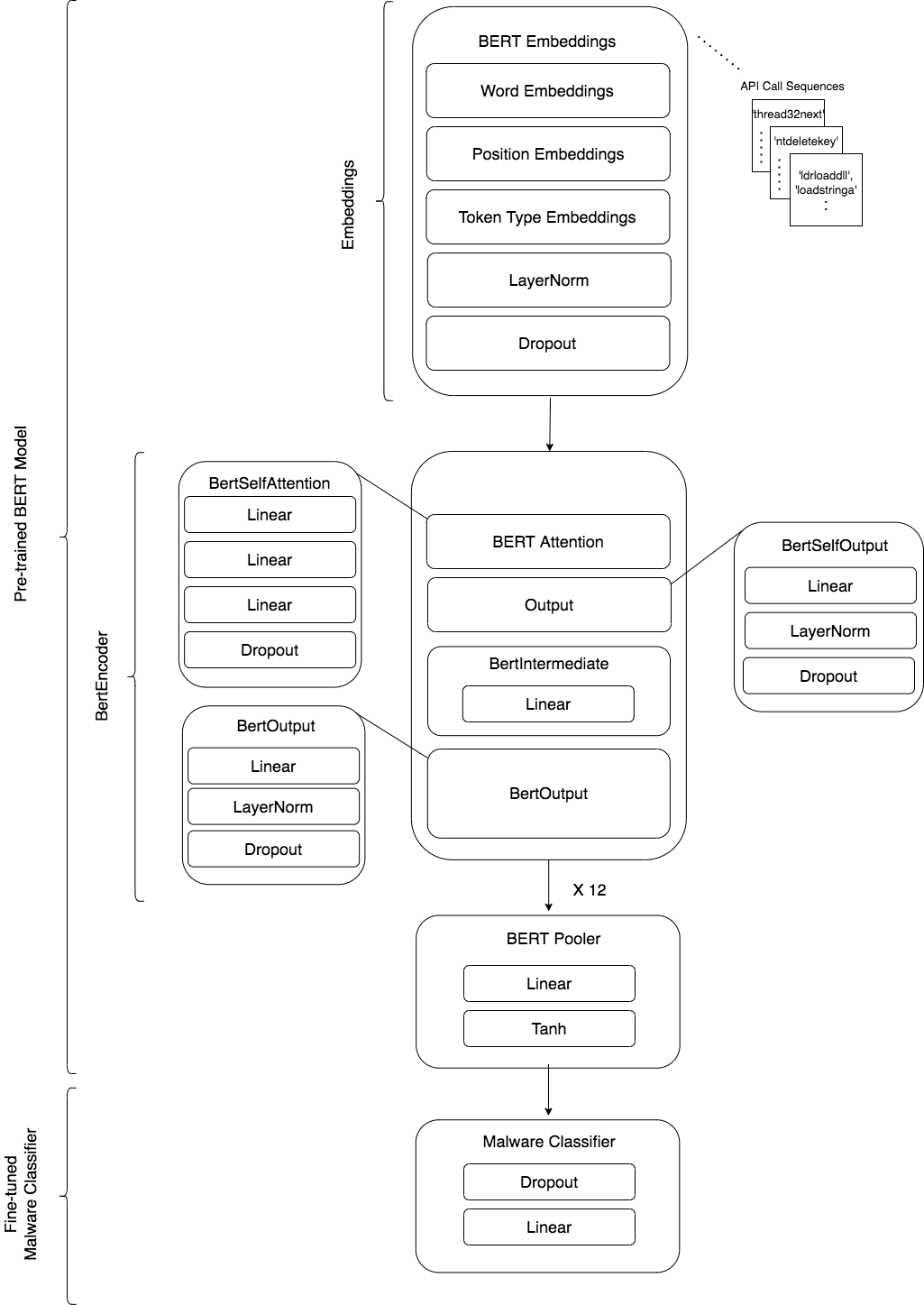}
\caption{BERT architecture.}
\label{fig:BertLasttt.png}
\end{figure}

BERT preserves the semantic content thanks to the masked language modeling (MLM) and next sentence prediction (NSP) unsupervised tasks, which enables to generate deep bidirectional representations while pre-training. BERT uses Word-piece tokenization to create a token vocabulary that consists of learned representations of the words. 
Figure \ref{fig:BertLasttt.png} shows BERT architecture in details.

\subsubsection{CANINE}
\label{sec:Canine}

CANINE, Character Architecture with No tokenization
In Neural Encoders, is a tokenizer-free pre-trained encoder model that is designed to overcome the shortcomings of the tokenization process such as word-piece and sentence-piece tokenization \cite{clark2021canine}. For example, a pre-trained model that uses specific tokenization may not be convenient for specialized domains. In \cite{boukkouri2020characterbert}, it is shown that word-piece tokenization based pre-training strategy is not well-suited compared to character-piece when fine-tuned on medical data.

\begin{figure}[H]
\centering
\includegraphics[width=\textwidth,height=\textheight,keepaspectratio]{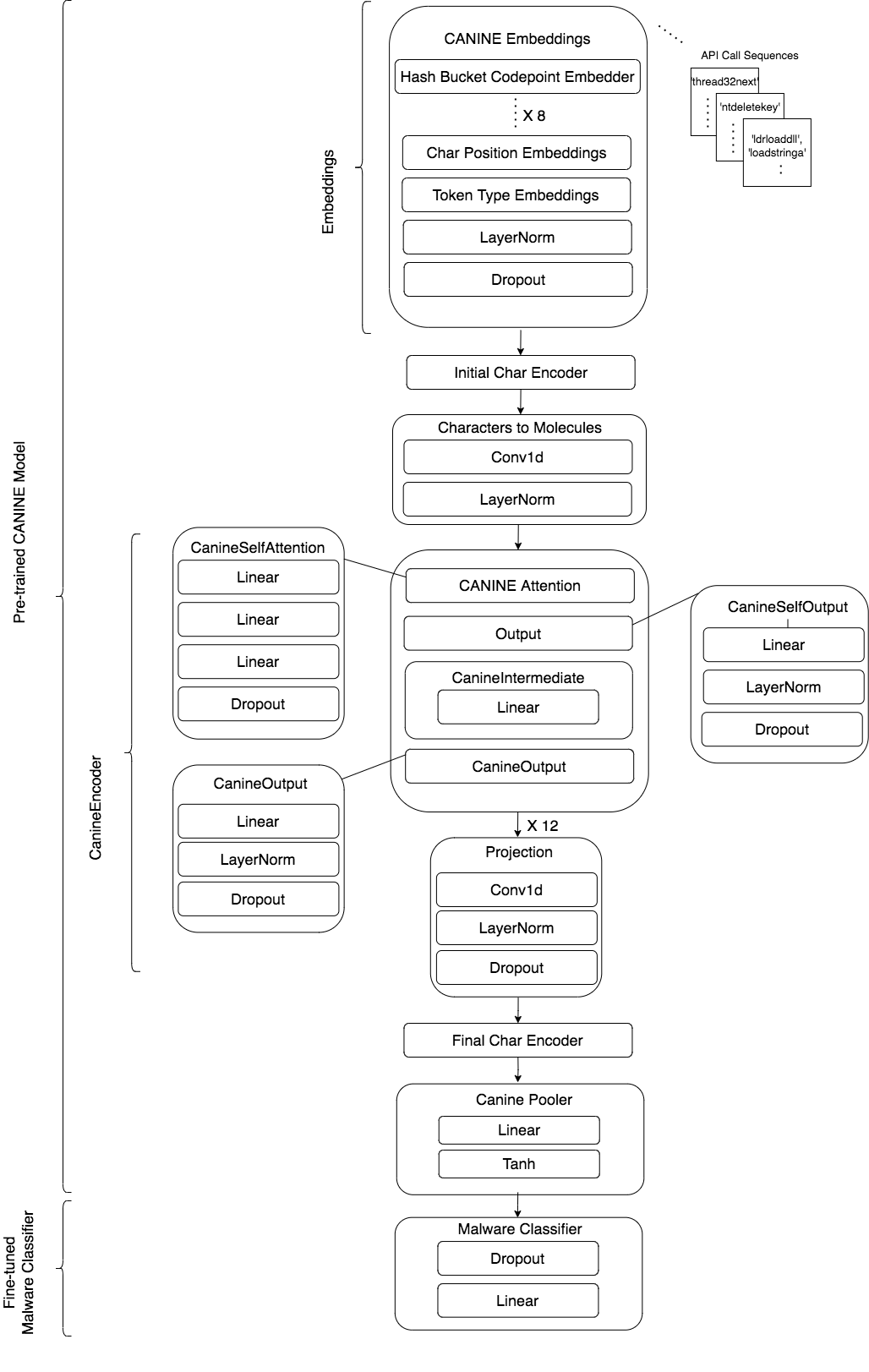}
\caption{CANINE architecture.}
\label{fig:Canine.png}
\end{figure}

Similar to BERT \cite{devlin2018bert}, CANINE is pre-trained on the MLM and NSP tasks as well.
Unlike commonly used pre-trained models, CANINE uses neural encoders that encode the sequence of characters or optionally sub-words are used as a soft inductive bias without doing explicit tokenization on input data. The CANINE structure is shown in Figure \ref{fig:Canine.png}.

In general, the BERT model is widely used by many studies for malware classification, as shown in the {\it Related Work}. We have assumed that the tokenization-free strategy used by the CANINE model might be well-suited for API calls since an API call such as 'ldrloaddll' may not be appropriate for word-tokenization. Therefore, we have included the CANINE model in our study. 

Thus, BERT, CANINE-C (Pre-trained with autoregressive character loss), and CANINE-S (Pre-trained with subword loss) pre-trained transformer models are leveraged regarding the \emph{RQ.4}. 

\subsection{Proposed Model: Random Transformer Forest (RTF)}

There are several important studies that show the success of using different ensemble types of pre-trained transformer models such as stacking and majority voting of heterogeneous pre-trained transformer models on varying downstream tasks. \cite{marcinczuk2021punctuation,morio2020hitachi,malla2021covid}. Unlike these type of ensemble models, Random Transformer Forest (RTF) is a bagging-based ensemble model inspired from the Random Forest (RF) machine learning model \cite{breiman2001random}. Similar to RF, using an ensemble of pre-trained transformer models is assumed to increase classification performance on highly imbalanced malware datasets rather than using a single pre-trained transformer model \cite{ccayir2021random,kobayashi2021reversed}.

The training phase requires creating \emph{N} different training subsets by using the bootstrap sampling method from the original training set. The malware class distribution coming from the original training set must be preserved in the resampling step due to the highly imbalanced class distribution. After the resampling step, each training subset is used to fine-tune the pre-trained transformer model. Each pre-trained transformer model has the same structure. Therefore base estimators are homogeneous.

\begin{figure}[ht]
\centering
\includegraphics[width=\textwidth]{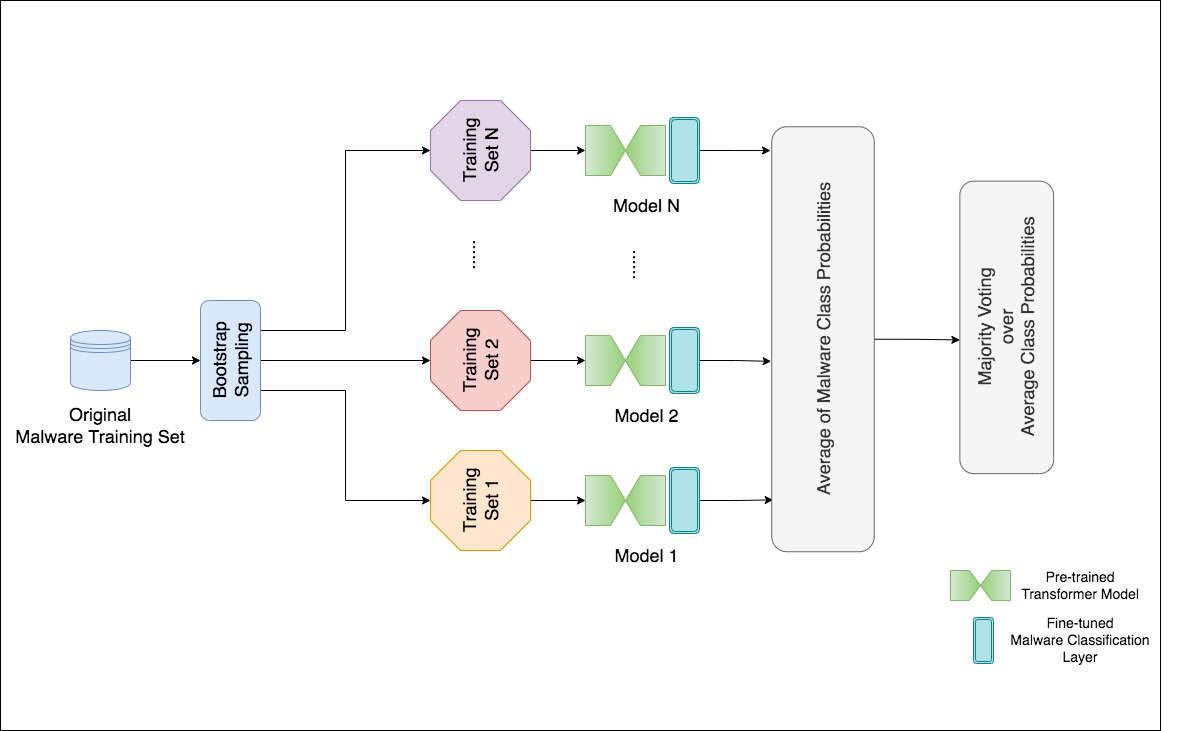}
\caption{The proposed RTF model architecture.}
\label{fig:RTFLast.png}
\end{figure}

In the testing phase, each fine-tuned transformer model takes a given malware API call sequence, and the class probabilities of each fine-tuned transformer model are aggregated by taking the average. For the majority voting, the final prediction is accepted as the malware family, which takes the highest probability coming from the aggregation part. 
Figure \ref{fig:RTFLast.png} shows the structure of RTF model. 


\section{Experiment and Results}
\label{sec:exp_res}

Experimental setup and conducted experiments will be clarified respectively regarding the predetermined Research Questions except \emph{RQ.1} which is elaborated in \emph{Methodology} part. 
\subsection{Experimental Setup}

We have utilized the google-cloud colab pro+ for our experiments. We have worked on Tesla T4 GPU with 51 GB available RAM provided by the colab platform. The Keras  framework \cite{chollet2018keras} is used for base model comparison. The PyTorch framework \cite{paszke2017automatic} is leveraged with pre-trained models taken from HuggingFace \cite{wolf2019huggingface} for pre-trained transformer models and RTF model. 
All jupyter notebook files that contain source codes regarding the research questions, from base model comparison to RTF model, can be found in the Github repository \footnote{\texttt{https://github.com/Ferhat94/Random-Transformer-Forest}}. 

\subsection{RQ.2-) What are the appropriate base models for multiclass malware classification based on API call sequences?}

In this part of the study, Transformer and LSTM model are leveraged as base architectures for comparison on four datasets. 


All datasets used to evaluate the base model performances are divided into three parts, training, validation, and testing. 20\% of the original datasets are allocated for testing. 
The splitting data process is performed in a stratified way as to preserve class distribution is one of the crucial steps on highly imbalanced datasets.
Considering the imbalanced distribution of the classes, we have leveraged the class weight approach to give different weights to both the majority and minority classes so that class weights are taken into account by training algorithms.

Stratified 10 Fold strategy is used on training data for each dataset, and 10\% of training data is used for validation for each iteration. 
Thus, we guarantee that each fold has the same distribution of malware families and ensure that every sample from the dataset has the chance of appearing in both training and validation data. 
Standard deviation and mean of 10 validation results are calculated for each evaluation metric used in our study, and training runtime for robust interpretation.

We have provided a dummy classifier as a simple baseline since class distribution of each dataset is imbalanced. We have used the "most frequent" strategy for our dummy classifiers. Base model comparison results on each dataset is shown in the \cref{table:baseOliveira,table:baseCatak,table:baseVirusSample,table:baseVirusShare}.




\begin{table}[H]
\centering
\caption{Base model comparison results for Catak \cite{catak2019benchmark} dataset.}
\begin{tabular}{@{}cccc@{}}
\toprule
\multicolumn{1}{c}{\textbf{Base Model}} & \textbf{F1-score} & \textbf{AUC score} & \textbf{Training Time (s)} \\ \midrule
\multicolumn{4}{c}{Validation Mean Scores} \\ \midrule
LSTM                 & 0.4873 $\pm$ 0.0126         & 0.7887 $\pm$ 0.0128     & 68.62 $\pm$ 11.44   \\ 
Transformer          & \textbf{0.5689 $\pm$ 0.0578}         & \textbf{0.8676 $\pm$ 0.0329}   & \textbf{11.74 $\pm$ 3.43}      \\\midrule
\multicolumn{4}{c}{Test Scores} \\ \midrule
LSTM                         & 0.4638         & 0.7885               \\
Transformer                                      & \textbf{0.5042}   & \textbf{0.8246}      \\ 
Dummy                      & 0.0308    & 0.5000   \\ \midrule
\end{tabular}
\label{table:baseCatak}
\end{table} 

\begin{table}[H]
\centering
\caption{Base model comparison results for Oliveira \cite{oliveira2019behavioral} dataset.}
\begin{tabular}{@{}cccc@{}}
\toprule
\multicolumn{1}{c}{\textbf{Base Model}} & \textbf{F1-score} & \textbf{AUC score} & \textbf{Training Time (s)} \\ \midrule
\multicolumn{4}{c}{Validation Mean Scores} \\ \midrule
LSTM                     & 0.5570 $\pm$ 0.0182       & 0.8853 $\pm$ 0.0142          & 111.99 $\pm$ 20.11          \\
Transformer              & \textbf{0.5792 $\pm$ 0.0379}        & \textbf{0.9280 $\pm$ 0.0142}      & \textbf{33.48 $\pm$ 7.91}             \\\midrule
\multicolumn{4}{c}{Test Scores} \\ \midrule
LSTM                         & 0.5637         & 0.8844               \\
Transformer             & \textbf{0.5650}   & \textbf{0.8855}      \\ 
Dummy                      & 0.0980    & 0.5000   \\ \midrule
\end{tabular}
\label{table:baseOliveira}
\end{table} 

\begin{table}[H]
\centering
\caption{Base model comparison results for VirusSample \cite{duzgun2021new} dataset.}
\begin{tabular}{@{}cccc@{}}
\toprule
\multicolumn{1}{c}{\textbf{Base Model}} & \textbf{F1-score} & \textbf{AUC score} & \textbf{Training Time (s)}\\ \midrule
\multicolumn{4}{c}{Validation Mean Scores} \\ \midrule
LSTM               & 0.7690 $\pm$ 0.0419    & 0.9656 $\pm$ 0.0110       & 21.80 $\pm$ 4.04             \\
Transformer           & \textbf{0.8070 $\pm$ 0.0323}     & \textbf{0.9885 $\pm$ 0.0055}  & \textbf{13.26 $\pm$ 3.38}                \\\midrule
\multicolumn{4}{c}{Test Scores} \\ \midrule
LSTM                         & 0.7531         & \textbf{0.9701}               \\
Transformer                                      & \textbf{0.7548}   & 0.9680      \\ 
Dummy                      & 0.1291    & 0.5000   \\ \midrule
\end{tabular}
\label{table:baseVirusSample}
\end{table} 

\begin{table}[H]
\centering
\caption{Base model comparison results for VirusShare \cite{duzgun2021new} dataset.}
\begin{tabular}{@{}cccc@{}}
\toprule
\multicolumn{1}{c}{\textbf{Base Model}} & \textbf{F1-score} & \textbf{AUC score} & \textbf{Training Time (s)}\\ \midrule
\multicolumn{4}{c}{Validation Mean Scores} \\ \midrule
LSTM                    & 0.7121 $\pm$ 0.0231         & 0.9274 $\pm$ 0.0130   & 96.79 $\pm$ 8.38                 \\
Transformer             & \textbf{0.7641 $\pm$ 0.0297}   & \textbf{0.9700 $\pm$ 0.0182}  & \textbf{31.76 $\pm$ 9.80}      \\ \midrule
\multicolumn{4}{c}{Test Scores} \\ \midrule
LSTM                         & 0.7071         & 0.9298               \\
Transformer                                      & \textbf{0.7125}   & \textbf{0.9350}      \\ 
Dummy                      & 0.0980    & 0.5000   \\ \midrule
\end{tabular}
\label{table:baseVirusShare}
\end{table}

Considering the two base models, LSTM and Transformer model, the standard deviation of the mean validation scores of evaluation metrics for the Transformer model is higher. Even in this case, we get higher results on unseen test data.

Evaluation results demonstrate that the Transformer model is more reasonable to continue with compared to the LSTM model in terms of evaluation results and training time.



\subsection{RQ.3-) What are the effects of pre-processing on API call sequences to the model results?}

Data pre-processing steps mentioned in {\it Pre-processing Method on datasets} part are applied to the Catak dataset \cite{catak2019benchmark} only as pre-processing has no effect on VirusSample and VirusShare datasets given in the study \cite{duzgun2021new} and Oliveira dataset \cite{oliveira2019behavioral} has already been pre-processed and limited with 100 non-consecutive API calls. 

Original API call sequences and pre-processed API call sequences on Catak dataset have been compared with the LSTM and Transformer model. 

Table \ref{table:preComparison} shows the comparison results of Original API call sequences and pre-processed API call sequences.

\begin{table}[H]
\centering
\caption{Comparison of the original and pre-processed Catak datasets.}
\begin{tabular}{@{}cccc@{}}
\toprule
\multicolumn{1}{c}{\textbf{Base Model}} & \textbf{F1-score} & \textbf{AUC score} & \textbf{Training Time (s)}\\ \midrule
\multicolumn{4}{c}{On Original API Call Sequences} \\ \midrule
LSTM  & 0.4638   & 0.7885 & \textbf{68.62 $\pm$ 11.44} \\
Transformer      & 0.5042   & 0.8246  & \textbf{11.74 $\pm$ 3.43}                  \\ \midrule
\multicolumn{4}{c}{On Pre-processed API Call Sequences} \\ \midrule
LSTM                                       & \textbf{0.5020}   & \textbf{0.8156} & 71.58 $\pm$ 12.27\\
Transformer                                     & \textbf{0.5106}   & \textbf{0.8372} & 18.27 $\pm$ 8.74       \\ \midrule
\end{tabular}
\label{table:preComparison}
\end{table}

Evaluation results demonstrate that pre-processing step outperforms for Catak dataset. Both the AUC score and F1-score have increased after the pre-processing steps for both LSTM and Transformer model. Although we expect training runtime to decrease after pre-processing steps, training runtime on the original Catak dataset is lower as the monitored evaluation metric, validation AUC, has stopped improving during the learning phase. Therefore, both LSTM and Transformer models have started to learn after pre-processing steps on the Catak dataset. New sequences generated after pre-processing steps are leveraged for the following experiments, pre-trained models and RTF model for the Catak dataset.


\subsection{RQ.4-) What are the effects of tokenizer-based (word piece) pre-trained transformer model (e.g. BERT) and tokenizer-free transformer model (e.g. CANINE) to our model results?}

In this part, Due to the large number of parameters used in pre-trained models, 20\% of the training data is allocated for validation instead of the stratified k fold strategy. The CANINE model uses two different pre-training strategies: sub-word loss (Canine-s) and character loss (Canine-c). In our studies both pre-training strategies are leveraged to see which pre-training strategy performs better for each dataset. Only the best Canine model, Canine-c or Canine-s,  is included to our final results from  \cref{table:rtfCatak,table:rtfOliveira,table:rtfVirusSample,table:rtfVirusShare} for each individual dataset.


The comparison of the pre-trained models, BERT and CANINE,  is shown at the end of the experiments with the RTF model results to see the comparison clearly.

\subsection{RQ.5-) What is the effect of ensemble of pre-trained transformer models, BERT and CANINE, which is based on bagging for imbalanced multiclass malware classification?}

In the RTF model \emph{N} different training subsets, thus \emph{N} different base estimators are utilized to fine-tune the \emph{N} different pre-trained BERT or CANINE model. 
We have tried several combinations of \emph{N} and pre-trained transformer models, BERT and CANINE, for each dataset. 
As a result of several trials, the combinations that provided the best scores are accepted as our RTF score.
Table \ref{table:bestParameters} shows the best combination for each dataset and model comparison results on each dataset are shown in \cref{table:rtfCatak,table:rtfOliveira,table:rtfVirusSample,table:rtfVirusShare}.

\begin{table}[H]
\centering
\caption{Best parameters for RTF model.}
\begin{tabular}{@{}ccc@{}}
\toprule
\multicolumn{1}{c}{\textbf{Dataset}} & \textbf{Number of Base Estimators (N)} & \textbf{Pre-trained Model} \\ \midrule
Catak \cite{catak2019benchmark}                    & 6   & BERT                     \\
Oliveira \cite{oliveira2019behavioral}                                 & 2    & BERT                    \\
VirusSample \cite{duzgun2021new}                                & 10      & CANINE-S                   \\ 
VirusShare \cite{duzgun2021new}                               & 5      & CANINE-S\\ \midrule
\end{tabular}
\label{table:bestParameters}
\end{table}


\begin{table}[H]
\centering
\caption{All model comparison on Catak \cite{catak2019benchmark} dataset.}
\begin{tabular}{@{}ccc@{}}
\toprule
\multicolumn{1}{c}{\textbf{Model}} & \textbf{F1-score} & \textbf{AUC score} \\ \midrule
Transformer                    & 0.5106   & 0.8372                     \\
CANINE-S                                  & 0.5633    & 0.8339                    \\
BERT                                 & 0.5919      & 0.8735                   \\ 
RTF                                  & \textbf{0.6149}      & \textbf{0.8818}                  \\ \midrule
\end{tabular}
\label{table:rtfCatak}
\end{table}


\begin{table}[H]
\centering
\caption{All model comparison on Oliveira \cite{oliveira2019behavioral} dataset.}
\begin{tabular}{@{}ccc@{}}
\toprule
\multicolumn{1}{c}{\textbf{Model}} & \textbf{F1-score} & \textbf{AUC score} \\ \midrule
Transformer                    & \textbf{0.5650}   & \textbf{0.8855}                     \\
CANINE-S                                  & 0.4725    & 0.8636                    \\
BERT                                 & 0.4839      & 0.8321                   \\ 
RTF                                  & 0.4061      & 0.8714                   \\ \midrule
\end{tabular}
\label{table:rtfOliveira}
\end{table}

\begin{table}[H]
\centering
\caption{All model comparison on VirusSample \cite{duzgun2021new} dataset.}
\begin{tabular}{@{}ccc@{}}
\toprule
\multicolumn{1}{c}{\textbf{Model}} & \textbf{F1-score} & \textbf{AUC score} \\ \midrule
Transformer                    & 0.7548   & 0.9680                     \\
CANINE-C                                 & 0.7893    & 0.9570                    \\
BERT                                 & 0.7759      & 0.9690                   \\ 
RTF                                  & \textbf{0.8059}      & \textbf{0.9773}                  \\ \midrule
\end{tabular}
\label{table:rtfVirusSample}
\end{table}

\begin{table}[H]
\centering
\caption{All model comparison on VirusShare \cite{duzgun2021new} dataset.}
\begin{tabular}{@{}ccc@{}}
\toprule
\multicolumn{1}{c}{\textbf{Model}} & \textbf{F1-score} & \textbf{AUC score} \\ \midrule
Transformer                    & 0.7125   & 0.9350                     \\
CANINE-S                                  & 0.7064    & 0.9286                    \\
BERT                                 & 0.7145      & 0.9364                   \\ 
RTF                                  & \textbf{0.7275}      & \textbf{0.9513}                   \\ \midrule
\end{tabular}
\label{table:rtfVirusShare}
\end{table}

According to the results, at least one of the pre-trained transformer models, CANINE or BERT, surpassed Transformer model, and the RTF model obtained the highest scores on three out of four datasets. 

\begin{figure}[H]
     \centering
     \begin{subfigure}[b]{0.49\textwidth}
         \centering
         \includegraphics[width=\textwidth]{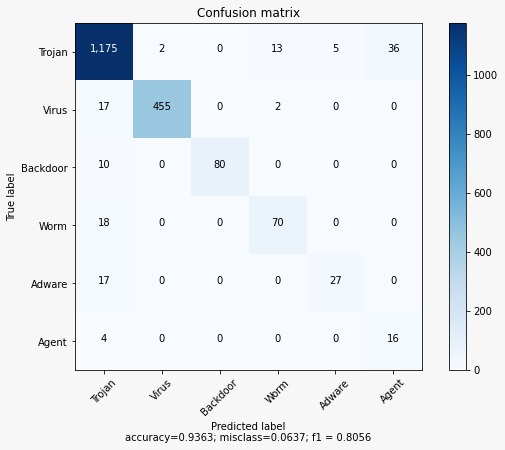}
         \caption{VirusSample CM}
         \label{fig:VirusSampleCM}
     \end{subfigure}
     \begin{subfigure}[b]{0.494\textwidth}
         \centering
         \includegraphics[width=\textwidth]{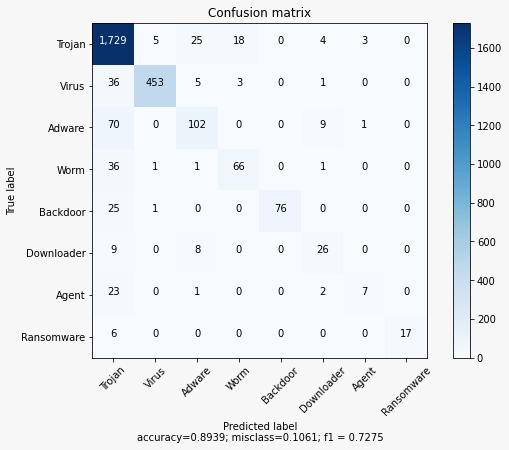}
         \caption{VirusShare CM}
         \label{fig:VirusShareCM}
     \end{subfigure}
     \newline
     \begin{subfigure}[b]{0.50\textwidth}
         \centering
         \includegraphics[width=\textwidth]{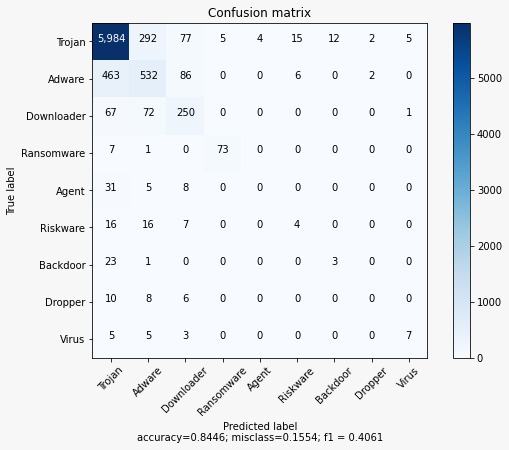}
         \caption{Oliveria CM}
         \label{fig:OliveiraCM}
     \end{subfigure}
        \caption{Performance results of RTF on each dataset for imbalanced classes}
        \label{fig:ThreeDatasetsCm.png}
\end{figure}

\begin{figure}[ht]
\begin{subfigure}{.5\textwidth}
  \centering
  \includegraphics[width=\linewidth]{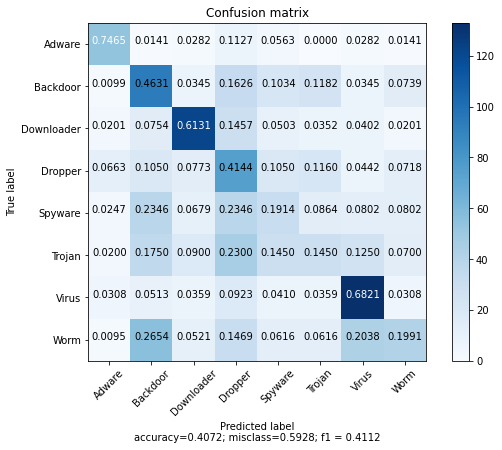}  
  \caption{SOURCE CM}
  \label{fig:sub-first}
\end{subfigure}
\begin{subfigure}{.5\textwidth}
  \centering
  \includegraphics[width=\linewidth]{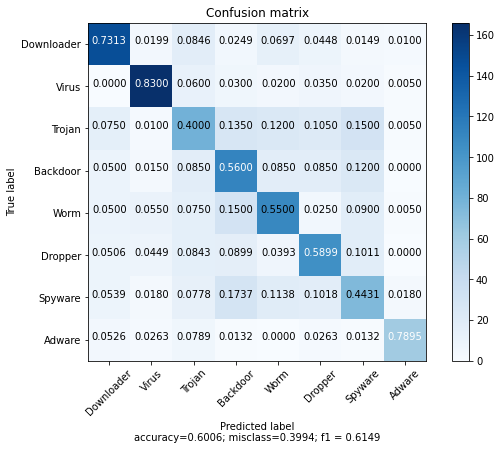}
  \caption{RTF CM}
  \label{fig:sub-second}
\end{subfigure}
\caption{Comparison of confusion matrix on Catak dataset.}
\label{fig:cmComparison.png}
\end{figure}

The performance of our proposed model RTF on VirusShare, VirusSample, and Oliveira dataset for each imbalanced class is shown in Figure \ref{fig:ThreeDatasetsCm.png}. The performance of the RTF on Catak dataset for each imbalanced class is shown in a separate figure as it is compared to the original authors' confusion matrix (CM) in Figure \ref{fig:sub-second} for the Catak dataset. 

Results show us that our proposed model is more accomplished in identifying minority classes on VirusSample and VirusShare dataset contrary to Oliveira dataset. Considering false negative predictions on these three dataset for each class, the model is prone to predict some minority classes as majority class which is Trojan in these cases. 


In our study, our proposed architecture (RTF) is compared with base models (LSTM and Transformer) and pre-trained transformer models (BERT and CANINE ) for each given dataset (Catak, Oliveira VirusSample, and VirusShare). Besides, since VirusShare and VirusSample datasets are newly published and Oliveira dataset is transformed to a multiclass problem by us, we have compared our approach with other approaches in the literature for the Catak dataset. 

In \cite{schofield2021comparison}, they have proposed a model for multiclass classification on Catak Dataset. 
In this study, all the samples in the Catak dataset are shuffled, and 80\% of the dataset is allocated for the train set. Then, all the samples in the Catak dataset are shuffled \emph{again} and 20\% of the dataset is allocated for the test set. The logical error made here is whole samples are shuffled twice. This situation causes the test part to have some samples which exactly fall into the train part. Thus, the model might test what it learned from the train. We have performed a test on the code shared with us by the authors \cite{schofield2021comparison} from the GitHub link \footnote{\texttt{https://github.com/MattScho/MalwareClassificationCNN}}. We have performed a test to divide the whole dataset into the training and testing dataset with the exact code script performed by authors for their study.
Finally, we realized that 1,117 of 1,371 samples allocated for the test set intersect with the train set. 
For these reasons, evaluation scores obtained by this article are not taken into account to compare our results. The logical error has been reported to the article authors.

To the best of our knowledge highest F1-score obtained on the Catak dataset for multiclass classification is 0.57 \cite{li2021api} compared to the baseline score of 0.47 obtained by the Catak dataset creators \cite{catak2020deep}. 
Although the F1-score reported by \cite{catak2020deep} is 0.47, the calculated F1-score from the given CM in \cite{catak2020deep} is 0.41 as in Figure \ref{fig:sub-first}. The 20\% of original Catak dataset is allocated as unseen test data for RTF experiments like in \cite{catak2020deep}.
In \cite{catak2020deep}, the authors showed their CM of LSTM model results on unseen test dataset. Their CM is referred to as source CM as this is the first study performed on Catak dataset.
Since only this study contains CM,  we have compared their CM with our proposed RTF model CM on unseen test data. 

Among the experimental studies conducted on the Catak dataset for multiclass classification \cite{li2021api,mcdonnell2021cyberbert,catak2020deep} our proposed RTF model has surpassed and reached the state-of-the-art F1-score of 0.6149 as shown in Figure \ref{fig:sub-second}. 

In our study, model performances are assessed according to F1-score and AUC score for pre-trained models, BERT, and CANINE, and RTF model thus far. However, besides F1-score and AUC score, training time is one of the factors to evaluate model efficiency. Therefore, training runtime results for BERT, CANINE, and RTF are given in Table \ref{table:trainingTime}.

\begin{table}[ht]
\centering
\caption{Training time (min.) comparison of BERT, CANINE, and RTF.}
\begin{tabular}{@{}ccccc@{}}
\toprule
\multicolumn{1}{c}{\textbf{Model}} & \textbf{Oliveira } & \textbf{VirusShare } & \textbf{Catak } & \textbf{VirusSample } \\ \midrule
BERT                                  & 141.02   & 26.25	& 28.63          & 18.73			      \\
CANINE                               & 58.27         & 11.12	& 12.03          & 7.77		                    \\
RTF 		         & 145.19  & 11.38	& 27.10          & 8.03		              \\ \midrule   
\end{tabular}
\label{table:trainingTime}
\end{table}

As seen in the Table \ref{table:trainingTime}, training runtime for pre-trained transformer models and our proposed model has increased dramatically compared to LSTM and Transformer model (\cref{table:baseOliveira,table:baseCatak,table:baseVirusSample,table:baseVirusShare}). The reason of high training runtime is BERT, and CANINE models are pre-trained on huge text data. 

Considering the security aspects of our study, higher training time does not hurt reaction time as these models will be trained first by AV scanners. 
Therefore, the inference time spent by AV scanners to process unseen data and make a prediction is the essential concern for security researchers to take an action as soon as possible. Inference time for each model is given in Table \ref{table:InferenceTime}.

\begin{table}[ht]
\centering
\caption{Inference time (s) comparison of each model.}
\begin{tabular}{@{}ccccc@{}}
\toprule
\multicolumn{1}{c}{\textbf{Model}} & \textbf{Oliveira } & \textbf{VirusShare } & \textbf{Catak } & \textbf{VirusSample } \\ \midrule
LSTM                                  & 0.3130   & 0.1110	& 0.3710          & 0.1070			      \\
Transformer                               & 0.0944         & 0.0931	& 0.0945          & 0.0924		                    \\
BERT 		         & 4.5137  & 2.7555	& 4.9658          & 1.4297		              \\CANINE                                  & 4.2778   & 2.6898	& 4.4324          & 1.4224			      \\RTF                                  &  4.5316  & 2.6856	& 4.9524          & 1.4183			      \\\midrule   
\end{tabular}
\label{table:InferenceTime}
\end{table}

Inference time given in Table \ref{table:InferenceTime}, shows us the processing and prediction time together for single unseen observation. 

\subsection{Limitations}
\label{sec:limitations}

As described in Table \ref{table:classDistribution}, even the four datasets used in our study consist of different malware families, we have total of 11 unique malware families. We think having a total of 11 different malware families is inadequate to understand malicious behavioural characteristics of a malware sample comprehensively. According to \cite{gregio2015toward}, malware taxonomy is divided into four different categories as stealing, evasion, disruption, and modification as these categories represent main malware behaviours. Therefore, among the defined behavior types such as Trojan, Virus, and Ransomware, several types need to be expanded and refined.

\section{Conclusion and Future Work}
\label{sec:conclusion}

In this study, we have leveraged several deep learning models for highly imbalanced multiclass malware classification based on API calls, which are inherently sequence problems. We have assessed the performance of our models with AUC score and F1-score evaluation metrics as the four datasets used in this study are imbalanced.  

Our evaluation results demonstrate that the Transformer model with one transformer block layer has achieved slightly better results than the LSTM model. Moreover, the pre-trained transformer models, BERT or CANINE,  outperformed one transformer block layer Transformer architecture. On the other hand, Transformer and LSTM models are noticeably faster than pre-trained models, BERT, and CANINE, in both training and inference times. However, considering the fact that training time does not directly affect the response time, and there are differences in the inference time on the basis of seconds, RTF model is more reasonable to continue with.

The CANINE model has been used for the first time in 
the field of malware classification in this study.
We have reached a state-of-the-art results on the static API calls datasets, VirusShare and VirusSample, with a bagging-based ensemble of the CANINE model. Therefore, we have demonstrated the success of the CANINE model with the strength of RTF.

We have achieved a state-of-the-art F1-score of 0.6149 on the Catak dataset with the power of bagging-based ensemble of BERT model and pre-processing since pre-processing steps have enabled us to increase our results significantly on the Catak dataset which has been built with dynamic API call sequences. In addition, the F1-score of 0.6149 obtained on the well-known benchmark Catak dataset has proved the success of our proposed RTF Model. 
In general, our proposed RTF Model has obtained state-of-the-art results on three out of four datasets.

This study can be extended by integrating our proposed ensemble model with the AUC maximization paradigm \cite{yuan2021large}. We believe we may increase our results in this way.

 \bibliographystyle{elsarticle-num} 
 \bibliography{cas-refs}





\end{document}